\documentclass[aps,preprintnumbers,amsmath,amssymb,prd,superscriptaddress,longbibliography,twocolumn]{revtex4-2}
\usepackage{dcolumn}
\usepackage{color}
\usepackage[caption=false]{subfig}
\usepackage{feynmp}
\usepackage{feynmp-auto}
\usepackage{systeme,mathtools}
\usepackage{stmaryrd}
\usepackage{easyReview}

\usepackage{mathtools}
\usepackage{nicefrac}

\def\comment#1{}

\newcommand{\nc}{\newcommand}
\nc{\scs}{\scriptstyle}
\nc{\setval}{\fmfset{wiggly_len}{3mm} \fmfset{arrow_len}{1.5mm}
	\fmfset{arrow_ang}{13} \fmfset{dash_len}{1.5mm}\fmfpen{0.125mm}
	\fmfset{dot_size}{2thick}}

\usepackage{bm,latexsym,mathrsfs,enumerate,color}
\usepackage[mathcal]{euscript}
\usepackage[breaklinks=true,unicode=true,urlcolor = blue,colorlinks = true,citecolor = blue,linkcolor = blue]{hyperref}
\usepackage{graphicx}
\usepackage{todonotes}
\usepackage{wrapfig}
\usepackage{easyReview}
\usepackage{physics}
\usepackage{amsmath}
\DeclareMathOperator{\sgn}{sgn}

\renewcommand{\vec}[1]{\bm{#1}}

\def\slashchar#1{\setbox0=\hbox{$#1$}           
	\dimen0=\wd0                                 
	\setbox1=\hbox{/} \dimen1=\wd1               
	\ifdim\dimen0>\dimen1                        
	\rlap{\hbox to \dimen0{\hfil/\hfil}}      
	#1                                        
	\else                                        
	\rlap{\hbox to \dimen1{\hfil$#1$\hfil}}   
	/                                         
	\fi}                                         %

\DeclareMathAlphabet\mathbfcal{OMS}{cmsy}{b}{n}

\begin{document}

\title{Superselection Rules, Bosonization Duality in 1+1 Dimensions and Momentum-Space Entanglement}

\author{Matheus H. Martins Costa}
\affiliation{Institute for Theoretical Solid State Physics, IFW Dresden, Helmholtzstr. 20, 01069 Dresden, Germany}

\author{Flavio S. Nogueira}
\affiliation{Institute for Theoretical Solid State Physics, IFW Dresden, Helmholtzstr. 20, 01069 Dresden, Germany}

\author{Jeroen van den Brink}
\affiliation{Institute for Theoretical Solid State Physics, IFW Dresden, Helmholtzstr. 20, 01069 Dresden, Germany}
\affiliation{Institute for Theoretical Physics and W\"urzburg-Dresden Cluster of Excellence ct.qmat, TU Dresden, 01069 Dresden, Germany}

\begin{abstract}
We investigate the effects of the presence of conserved charges on the momentum-space entanglement of Quantum Field Theories (QFTs). We show that if a given model has superselection sectors, then it allows for different notions of momentum modes, each associated with a complete set of commuting observables. Applying this idea to investigate the entanglement of momentum degrees of freedom on both sides of the abelian Bosonization duality, we find that different tensor product partitions are mapped into each other and give explicit examples to sustain our findings. The conditions on which our conclusions may be generalized to other duality transformations, which require the introduction of a notion stricter than the most general possible, are laid and directions for further work are given.
\end{abstract}

\maketitle

\section{Introduction}

Dualities between different quantum field theories (QFTs) offer key insights when studying their structure and phase diagram \cite{Savit_RevModPhys.52.453,BANKS1977493,Peskin1978,THOMAS1978513,Dasgupta-Halperin_PhysRevLett.47.1556,SEIBERG199419,SEIBERG2016395,CARDY198217,CARDY19821,Nogueira-Nussinov-van_den_Brink_PhysRevD.94.085003,Cobanera_PhysRevLett.104.020402,bondalgebra}, a powerful example being the bosonization duality, with its more prominent examples being in 1+1 \cite{Coleman_PhysRevD.11.2088,gogolin2004bosonization,1dquant} and 2+1 dimensions \cite{SEIBERG2016395,Karch_PhysRevX.6.031043,Mross_PhysRevX.7.041016,Senthil_PhysRevX.7.031051,Raghu_PhysRevLett.120.016602,FERREIROS20181,NASTASE2018145,Nogueira_PhysRevD.102.034506,Shyta_PhysRevLett.127.045701,Shyta_PhysRevLett.129.227203,Shyta_PhysRevD.105.065019}. With the rising interest over the past decades in studying quantum information properties of field theories and many-body systems \cite{nishioka, PhysRevLett.90.227902}, the question naturally arises as to whether or not the entanglement entropy is invariant under these transformations. 

As shown in Ref. \cite{radivcevic2016entanglement}, this is a subtle matter which depends on what is meant by ``invariant": from the point of view of the algebraic formalism \cite{entropy, haag}, one obtains that the ground state entanglement entropy of the ``bond"  algebras of observables mapped into each other by the Wannier duality are the same. However, if instead the question asked is whether the entanglement entropy of a given region of space (here understood in terms of tracing out the degrees of freedom of the complementary region, or equivalently, choosing an appropriately large local algebra of observables) is the same on both sides of the duality, Ref. \cite{radivcevic2016entanglement} also demonstrated that for lattice systems this is generally not the case. 

At the same time, considering the Bosonization duality in 1+1 dimensions in the continuum, it was shown via an explicit calculation in \cite{Headrick_2013} that, at least for the case where the duality maps free fermions into free bosons, the full reduced density matrix (thus, in particular, the entanglement entropy) of a region of space is preserved, contrasting with the lattice spin results. Furthermore, this conclusion is expected to be valid for any point in the parameter space of the duality, leading to a use of the duality to calculate real-space entanglement measures in an interacting fermionic theory in Ref. \cite{PhysRevD.108.125016}. Such expectation is due to the fact that both fermionic and bosonic representations have the same scaling operators, a topic to which we will return later.

With these contrasting results in mind, and given the recent interest in studying entanglement of quantum field theories in momentum-space  and its connections to the renormalization group \cite{Balasubramanian:2011wt, Hsu:2012gk, paper1, PhysRevD.108.085004, paper2}, it is natural to ask whether or not momentum-space entanglement is preserved under duality transformations.  

The question posed is obviously very broad, so in this work we will focus on the fermion-boson duality in 1+1 dimensions in the continuum, as in the previously mentioned real space analysis in Ref. \cite{Headrick_2013}, and see which lessons can be applied generally. Due to the inherent presence of charged sectors on both sides of the duality \cite{swieca}, we are led to first investigating what changes in the momentum-space entanglement of a QFT occur in the presence of conserved charges and superselection sectors, in analogy to the real space studies of Refs. \cite{Casini:2019kex, Furuya:2020wxf, Casini:2020rgj}.

This paper will proceed as follows. Our first result in Section \ref{sec2} shows that QFTs with superselection sectors, such as fermionic theories, naturally allow for multiple notions of partitioning the Hilbert space into momentum degrees of freedom, none more ``fundamental" than the other. This lays the groundwork to understanding what may happen under the Bosonization transformation, as we conclude that certain dualities map the different tensor product structures into each other. Our results follow directly from the fact that for theories with superselection rules there are several different sets of operators which may be used to construct the QFTs observables, as investigated in Refs. \cite{currents1,currents2}. By taking an appropriate Fourier transform, this is what leads to different partitions into momentum modes. More generally, our result shows that every complete set of commuting observables determines such a tensor product structure and conserved charges allow for new sets of commuting operators to be complete within a sector.

In Section \ref{sec3} we illustrate the previous discussion concretely, using the case of the (abelian) Bosonization duality to show that a direct calculation of the ground state momentum-space entanglement entropy is not invariant, but different partitions remain preserved. We also perform an explicit calculation at lowest nontrivial order of the momentum-space entanglement entropy in the Sine-Gordon model (taking the momentum-space subsytems to be the set of spatial momenta with $|\Vec{k}|<\mu$ and $|\Vec{k}|\geq\mu$). From Section \ref{sec2} this entanglement entropy is also the entanglement between modes of the fermionic current in certain models, such as the massive Thirring \cite{coleman} and the Luttinger liquid \cite{Haldane_1981, schulz1998fermi}. In particular, at the ``free fermion point" of the Bosonization duality, the ground state is separable for one partition, while for the other it is non-trivially entangled.

By combining this new understanding of possible momentum partitions with the proof in Ref. \cite{paper2} of the momentum-scale separability of renormalization group (RG) fixed points, Section \ref{sec4} shows that regardless of the choice of tensor product, the fixed point ground state is still separable. This is explicitly discussed for the massless Thirring model.

We then conclude in Section \ref{sec5} with an investigation on the possible forms of duality transformations and their consequences for tensor product partitions in momentum-space. As shown in Ref. \cite{bondalgebra}, the most general duality maps only need to preserve certain structures of the Hamiltonian operator (the ``bond algebras") and are thus von Neumann algebra isomorphisms. From this we generalize the results in Ref. \cite{radivcevic2016entanglement} for real-space entanglement, leaving open the way for future studies of the analogous consequences in momentum-space. Finally, we introduce a stronger notion of duality, which we name a ``local duality", that preserves the full local algebras of observables, and show that these are precisely the transformations which map different momentum-space partitions into each other, as in the Bosonization case.

\section{Momentum-space entanglement and superselection sectors}
\label{sec2}

When discussing entanglement, the fundamental idea is that of restricting the set of observables to subsystems of the global state. Mathematically, this translates into the notion of taking the appropriate subalgebras of observables \cite{haag,witten}, which is equivalent to the usual partition of the Hilbert space into a tensor product in finite-dimensional systems. It is well-known that for QFTs one can no longer associate regions of space with tensor product factors \cite{witten, mutualinfo, buchholztype3, arakitype3}, but one can still think in these terms heuristically, as long as proper care is taken. The latter is the approach we will follow in this work, with the understanding that while a full algebraic definition of the momentum-space entanglement we are interested in does not yet exist, the ideas leading towards such construction underlie our thinking (see the comments in the conclusion of Ref. \cite{paper2}).

Our purpose with the preamble above is to recall the importance of focusing on the allowed/physical observables, a criterion which becomes especially relevant when discussing entanglement in fermionic systems or those with conserved global charges.

In a generic QFT, we may use the field operator $\phi(x)$ to construct local observables and its Fourier transform $\phi(k)$ to generate the operators associated with regions of momentum-space \cite{Balasubramanian:2011wt, paper1, paper2}. However, if the field is charged under a global symmetry, this is no longer a physical operation, as acting on a state with it will change the total charge number. The same happens to any function of $\phi(x)$ which is not invariant under the QFT's symmetry. Thus, the set of observables of the theory is restricted to those operators carrying zero charge, i.e., constructed from an equal number of $\phi$ and $\phi^\dagger$ operators.

What we have just described is the natural setup of field theories with superselection rules, for which different sectors cannot be connected via the application of physical operators, see Refs. \cite{www1, www2}. It turns out that the existence of different charged sectors has a strong influence on the entanglement properties of the ground state, a direct consequence of this restriction of the allowed observables to the set of only those which carry no charge.

For the entanglement of local regions of space, the consequences have been explored in great detail in Refs.\cite{Casini:2019kex, Furuya:2020wxf, Casini:2020rgj}, which we will review shortly in order to gain intuition as to how the presence of internal charges can affect entanglement. After this discussion, we will then focus on exploring how these subtleties manifest in momentum-space entanglement. In particular, we will show how superselection sectors allow for the definition of new inequivalent tensor product partitions of the Hilbert space of the QFT into momentum modes.

A final note before moving forward: our studies in this paper are restricted to QFTs with global symmetries (mathematically, those with ``DHR superselection sectors", where DHR refers to Haag, Doplicher, Roberts \cite{DHR-1,DHR-2,DHR-3}, see Refs.\cite{Casini:2019kex, Furuya:2020wxf, Casini:2020rgj} for an introduction), while Ref. \cite{Casini:2020rgj} goes further to include gauge theories and even those with the so-called "generalized symmetries" \cite{gensym, shao2024whatsundonetasilectures}. We stay with the simplest case, as momentum-space entanglement is less well-understood as its real space counterpart and the global symmetries already reveal interesting structures on their own.

\subsection{Review of the implications in real space}

In a QFT with the presence of conserved global charges, any quantum fluctuations of a state (in particular, of the ground state, which will be the case throughout this paper) are necessarily such that if charged excitations appear, they do so in groups of vanishing total net charge. The simplest case being, of course, pairs of particles having opposite charges appearing spontaneously.

Let us now partition the spatial degrees of freedom into three regions: $A$ and $B$ which are separated, plus their complement $C=(A\cup B)'$. When looking at the mutual information between $A$ and $B$, then, as discussed in Ref. \cite{Casini:2019kex}, it is possible for a fluctuation to create a charged particle inside $A$ and an anti-charged particle in $B$. This, then, raises a question: should such fluctuations be taken into account when calculating entanglement between $A$ and $B$?

The answer found in the works of Refs. \cite{Casini:2019kex, Furuya:2020wxf, Casini:2020rgj} is that both possibilities, including or excluding the fluctuations described above, are consistent and physically meaningful. In Ref. \cite{Casini:2020rgj} this is explained with the algebraic formalism, where there is an unavoidable ambiguity in the algebras of observables that may be associated with $A\cup B$. In this brief review, we will restrict ourselves to describing the physical mechanism that leads to this ambiguity and how it is related to the initial discussion on physical versus charged operators.

In more detail, by thinking in operational terms (how many Bell pairs can be distilled from the entanglement between $A$ and $B$, for example) there are two setups we may conceive of (see Sec. 2 of Ref. \cite{Furuya:2020wxf}): in the first we allow each ``lab" to act exclusively on $A$ or $B$, therefore, since only charge-neutral observables are physical, this does not allow for the use of an operator of the form $\phi^\dagger(x)\phi(y)$ and so the sort of particle-antiparticle fluctuation previously described cannot be used for quantum protocols. On the second setup, we permit such operations and their generalizations with the creation of multiple pairs (by including the "intertwiners" in the algebra of observables, see Refs. \cite{Casini:2019kex, Furuya:2020wxf}) and so reach the maximum potential for use of entanglement \cite{Furuya:2020wxf}. The price to pay, however, is that operators such as $\phi^\dagger(x)\phi(y)$ can only be physically generated non-locally \cite{Casini:2020rgj}, by first acting on the vacuum in some position $z$ with the physical observable $\phi^\dagger(z)\phi(z)$ to create a particle-antiparticle pair and then continuously moving the charges to their final positions at $x$ and $y$.

Each of the possibilities described has a clear physical interpretation and can be associated to an entanglement entropy \cite{Furuya:2020wxf, Casini:2020rgj}. The lack of any rule that might privilege one choice over the other is not a flaw in our formulation of the QFT, but rather an indication of how superselection rules enrich the entanglement structure of such systems. In fact, this connection between superselection sectors and quantum non-locality has been studied extensively, see Ref. \cite{PhysRevLett.91.010404}.

As we move on to the study of momentum-space entanglement, we must keep this in mind: by restricting operators to only the physical observables, the notions of entanglement in the system involve subtleties that do not exist in the absence of superselection rules. More specifically, the inclusion or not of operators which are generated non-locally changes the nature of the entanglement structure being considered. This will be relevant as we move forward.

\subsection{Partitions in momentum-space}

So when defining entanglement in the momentum space of a QFT, we cannot use the field operator $\phi(\Vec{k})$ by itself as it carries charge. However, drawing from the lessons learned in the real space scenario, we can still define a tensor product partition of the Hilbert space $\mathcal{H}$ of the theory which we would naively associate with $\phi(\Vec{k})$ by taking the set of observables generated by the charge-neutral combination $\phi^\dagger(\Vec{k})\phi(\Vec{k})$. In free theories, this is simply operators of the form $a^\dagger_{\Vec{k}}b^\dagger_{\Vec{k}}$, $a^\dagger_{\Vec{k}}a^\dagger_{\Vec{k}}$ and so on, plus their combinations, where $a^\dagger$, $b^\dagger$ are the particle and anti-particle creation operators, respectively. For interacting models, we must, of course, properly renormalize the composite operator.

That this construction indeed leads to a partition $\mathcal{H}=\bigotimes_{\Vec{k}}\mathcal{H}_{\Vec{k}}$ can be more easily seen via the path integral. In Refs. \cite{Balasubramanian:2011wt, paper1, paper2} the existence of a momentum-space tensor product is derived from the factorization of the path integral measure $\mathcal{D}\phi = \prod_{\Vec{k}}\mathcal{D}\phi_{\Vec{k}}$, which shows that there are subalgebras of observables associated with sets of momenta. Here, the presence of a charged field implies that for physical operators it is meaningless to integrate $\phi$ and $\phi^\dagger$ independently, however, one can still write $\mathcal{D}\phi^\dagger\mathcal{D}\phi = \prod_{\Vec{k}}\left(\mathcal{D}\phi^\dagger_{\Vec{k}}\mathcal{D}\phi_{\Vec{k}}\right)$ and so define subalgebras generated from bilinears of $\phi_{\Vec{k}}$. Furthermore, since all operators can be constructed as a function of $\phi_{\Vec{k}}$, we conclude that no physical observables are left out in this procedure, leading to a consistently-defined partition.

We will denote the tensor product thus defined by the expression $\mathcal{H}=\bigotimes_{\Vec{k}}\mathcal{H}_{\Vec{k}}^{\phi}$, where $\mathcal{H}_{\Vec{k}}^{\phi}$ indicates the Hilbert space of momentum $\Vec{k}$ degrees of freedom on which $\phi(\Vec{k})^\dagger\phi(\Vec{k})$ acts. Interestingly, in the context of entanglement in fermionic systems (which always contain a parity superselection rule due to the anticommutation relations), the work of Ref. \cite{entfermion} proved that when the global state is pure, the notion of separability of modes does not change if the charge-carrying operators are included as part of the allowed observables or not, see Sec. \ref{sec3} in particular. This makes for a good test of our assertion that we may use the path integral measure to identify valid tensor product partitions in the Hilbert space $\mathcal{H}$: when integrating out modes in order to take a partial trace of the ground state \cite{Balasubramanian:2011wt, paper1, paper2}, the path integral ``does not know" whether or not the we are including the charge-carrying field as a valid operator, and thus the notion of entanglement obtained this way must be indifferent to this choice, which is directly proven to happen in the fermionic case.

Now, with a momentum-space partition being directly derived from the field $\phi_{\Vec{k}}$ of QFT, how can other tensor product structures appear as we have claimed? The key is to look back at our discussion on physical observables which are local in real space. As we recalled, only the charge neutral operators are observable and the simplest of those are bilinears of the field. In fact, we may intuitively expect that all local observables can be written as functions of bilinears, in which case expressions of the form $:\phi^\dagger(x)\phi(x):$ now play the role of ``fundamental field". With this description of the degrees of freedom, one can expect that, in the same way $\phi_{\Vec{k}}$ as a function of $\Vec{k}$ is associated with certain tensor product factors in momentum-space, so can we define a partition of $\mathcal{H}$ in terms of $:\phi^\dagger\phi:(\Vec{k})$. Furthermore, we have, formally, the equality

\begin{equation}
\label{convolution}
    :\phi^\dagger\phi:(\Vec{k}) = \int\frac{d^dp}{(2\pi)^d}\phi^\dagger(\Vec{k}+\Vec{p})\phi(\Vec{p}),
\end{equation}
meaning that the partition thus defined is not the same as $\bigotimes_{\Vec{k}}\mathcal{H}_{\Vec{k}}^{\phi}$, since we are associating with a single mode $\Vec{k}$ field configurations which involve all the $\mathcal{H}_{\Vec{k}}^{\phi}$. This is the analogous of the use of non-local operators we discussed previously.

So far our considerations above have been heuristic, meant to give some intuition. Let us now make all the steps more precise. First and foremost, it is indeed possible to fully describe the degrees of freedom/algebra of observables of a QFT in the presence of superselection rules and this was proven in Refs. \cite{currents1, currents2}. The authors' main result is to formally show that in the presence of conserved charges, the charge densities and current components plus a finite number of charge neutral bilinears of the field (the exact form of which depend on whether the theory described is bosonic or fermionic, relativistic or not) define an irreducible set of operators within any individual superselection sector, i.e., every physical (charge-neutral) observable can be written as a function of these operators.

For illustration purposes, we reproduce the proof in Ref. \cite{currents1} of the fact that in a relativistic free fermionic theory, the field bilinears (which we will generically call ``currents") are sufficient to generate all observables. Proofs for bosonic and/or non-relativistic fields follow similarly. We will discuss afterwards our use of free fields.

Given the fermionic field at a fixed time $\psi(\vec{x})$, we define $J_M(\vec{x}) = :\Bar{\psi}(\vec{x})\hat{M}\psi(\vec{x}):$ for all matrices $\hat{M}$ acting on the space of spinors. Assuming there are no other quantum numbers besides spin, one can write $\hat{M}$ as a linear combination of the identity, gamma matrices $\gamma_{\mu}$, commutators $\sigma_{\mu\nu} = [\gamma_{\mu},\gamma_{\nu}]$ and so on, as usual. Generalization to other cases is simple.

We also introduce the ``current" operators given by $\Tilde{J}_i(\vec{x}) = \frac{1}{2i}:\left(\Bar{\psi}(\vec{x})\partial_{x_i}\psi(\vec{x})-\partial_{x_i}\Bar{\psi}(\vec{x})\psi(\vec{x})\right):$ which are components of, and can be replaced by, the spatial parts of the energy-momentum tensor \cite{currents1, currents2}. With this, dealing with the free field divergences in the usual way, the current commutators obey the identities,
\begin{align}
    \begin{split}
    \label{comm}
        \left[J_M(\vec{x}),J_{M'}(\vec{y})\right] &= \delta(\vec{x}-\vec{y})J_{[M,M']}(\vec{x})\\
        \left[J_M(\vec{x}),\Tilde{J}_i(\vec{y})\right] &= -i\partial_{x_i}\left(\delta(\vec{x}-\vec{y})J_M(\vec{x})\right) \\
        \left[\Tilde{J}_i(\vec{x}),\Tilde{J}_j(\vec{y})\right] &= -i\partial_{x_i}\left(\delta(\vec{x}-\vec{y})\Tilde{J}_i(\vec{x})\right) \\ &+i\partial_{y_j}\left(\delta(\vec{x}-\vec{y})\Tilde{J}_j(\vec{x})\right).
    \end{split}
\end{align}

Suppose now, $\mathbb{O} = \mathbb{O}(\psi,\psi^\dagger)$ is any observable commuting with all $J_M(\vec{x})$ and $\Tilde{J}_i(\vec{x})$. Since it commutes with $\rho(\vec{x}) = :\psi^\dagger(\vec{x})\psi(\vec{x}):$, it is invariant under the unitary $U(f) = \exp{i\int d^dxf(\vec{x})\rho(\vec{x})}$ for any function $f(\vec{x})$ (strictly speaking, $f(\vec{x})$ must be quickly decreasing in order to obtain a well-defined operator \cite{haag}, this is enough for our proof). 

From $[\psi(\vec{x}), \rho(\vec{y})] = \delta(\vec{x}-\vec{y})\psi(\vec{x})$ the unitaries act on the field as $U(f)\psi(\vec{x})U^\dagger(f) = e^{if(\vec{x})}\psi(\vec{x})$. Thus, the invariance of $\mathbb{O}$ imposes that it is a function of the form $\mathbb{O} = \mathbb{O}(:\psi^\dagger(\vec{x})\psi(\vec{x}):) = \mathbb{O}(\rho(\vec{x}))$, as can be seen from its formal Taylor expansion. 

This form already guarantees that $\mathbb{O}$ commutes with all $J_M(\vec{x})$, a consequence of Eq. (\ref{comm}). By commuting with the $\Tilde{J}_i(\vec{x})$,  we use standard functional calculus results to see that $\mathbb{O}$ must then satisfy
\begin{equation}
    \left[\Tilde{J}_i(\vec{x}), \mathbb{O}\right] = i\partial_{x_i}\rho(\vec{x})\frac{\delta\mathbb{O}}{\delta\rho(\vec{x})} = 0.
\end{equation}

Since this is valid at any point of space, the most general form of all such operators $\mathbb{O}$ is, given an arbitrary number $c$, necessarily,
\begin{equation}
    \mathbb{O}(\rho(\vec{x})) = c\int d^dx \rho(\vec{x}),
\end{equation}
being thus proportional to the total charge.

Finally, since by definition the physical observables do not change the superselection sector of the state, we conclude that any operator commuting with all currents is proportional to the identity within a sector. Thus any observable can be written as a function of these bilinears.

Before moving forward, note that in this proof (and in Refs. \cite{currents1, currents2} more generally) we have relied on the existence of certain equal-time commutators of the currents. For theories with a cutoff, such as in the lattice, this means that we may proceed with our discussion without further concerns, as the existence of the commutators used in previous paragraphs is guaranteed by the validity of the canonical (anti)commutation relations. However, for interacting QFTs in the continuum these will generally not be well-defined, because new terms are expected to appear and operators at a fixed time will not always exist as they become too singular \cite{haag}. In these latter cases, we still expect that the main point, the irreducibility of field bilinears in theories with conserved charges, will hold based on the proof for regularized theories. The results of the next Section, involving continuum relativistic QFTs, signal that this is a reasonable assumption and so for the remainder of this work we will treat it as true.\bigskip

With a full description of the physical degrees of freedom in terms of bilinears following from irreducibility, our next step is to take a portion of them which generates a complete set of commuting observables. Going back to our example of relativistic fermions and assuming there are no charges other than the fermion number (which does not change the discussion qualitatively), we may choose as generators $\rho(x)$ and $J_1(\vec{x}) = :\Bar{\psi}(\vec{x})\psi(\vec{x}):$. Since no other bilinear commutes with these, a basis for the full Hilbert space of the system is formally given by their eigenvectors, which we denote by $\ket{\rho(\vec{x}),\Bar{\psi}\psi(\vec{x})}$, and it allows us to write a path integral representation of the QFTs ground state.

This construction, called the ``density representation" (we may also build other ``current representations", see Refs. \cite{currents1,currents2}), has long been useful for the study of certain systems, see for example Ref. \cite{FRADKIN1993667}. What we point out is that the Fourier transforms $\rho(\Vec{k}), J_1(\Vec{k})$ also generate a complete set of commuting observables, meaning we can use their eigenvectors $\ket{\rho(\Vec{k}), \Bar{\psi}\psi(\Vec{k})}$ to identify Hilbert space factors $\mathcal{H}^\rho_{\Vec{k}}$ and thus the tensor product partition $\bigotimes_{\Vec{k}}\mathcal{H}_{\Vec{k}}^{\rho}$.

In detail, in the path integral formalism we can associate ``canonically conjugate momenta" to $\rho(\Vec{k}), J_1(\Vec{k})$ via their respective functional derivatives $\frac{\delta}{\delta\rho(\Vec{k})}, \frac{\delta}{\delta J_1(\Vec{k})}$ and build operators in the Schr\"odinger representation \cite{jackiw}. The factorization of the path integral measure in momentum modes implies the existence of the tensor product structure, just as in the field representation. 

Equivalently, we can also show this by taking the set of operators $\{\rho(\Vec{k}), J_1(\Vec{k}) | \Vec{k} \in \Tilde{\mathcal{O}}\}$ for any region of momentum space $\Tilde{\mathcal{O}}$ and constructing the Hilbert space $\mathcal{H}_{\Tilde{O}}$ from their eigenvectors $\ket{\rho(\Vec{k}), J_1(\Vec{k})}$. That these are, formally, tensor factors can be seen from the commutation of operators in $\mathcal{B}(\mathcal{H}_{\Tilde{O}})$ with those in $\mathcal{B}(\mathcal{H}_{\mathbb{R}^d \setminus \Tilde{O}})$, the sets of bounded operators acting on the spaces of momentum modes. Thus, again we arrive at the characterization of a different partition of the full Hilbert space.

That the tensor products in the density representation is distinct from the one previously considered in the ``field representation" $\bigotimes_{\Vec{k}}\mathcal{H}^\psi_{\Vec{k}}$ (with $\psi$ playing the role of $\phi$) can be easily seen from the fact that $\rho(\vec{k})$, for example, being the Fourier transform of a bilinear as in the example of Eq. (\ref{convolution}), acts on all modes we associate with $\psi(\vec{k})$. In particular, these operators do not commute at different momenta, so they are not associated to the same momentum-space subalgebras.

Such distinction between different partitions is not just a theoretical curiosity, it can be directly related to experimental concepts and procedures. In a condensed matter context, the density representation $\bigotimes_{\Vec{k}}\mathcal{H}^\rho_{\Vec{k}}$ is directly associated with the Fourier components of n-point density correlation functions $\langle \rho(\vec{x_1})\rho(\vec{x_2})...\rho(\vec{x_n})\rangle$, while the field representation $\bigotimes_{\Vec{k}}\mathcal{H}^\psi_{\Vec{k}}$ involves the expectation values of operators such as $\psi^\dagger(\Vec{k})\psi(\Vec{k})$, which can be directly measured via procedures such as ARPES (angle-resolved photoemission spectroscopy) \cite{arpes}. Hence, each partition leads to a different experimental setup, which is always the case when considering inequivalent subsystems.

We conclude this Section by commenting that for any maximal set of commuting observables a momentum-space partition may be defined, and vice-versa. The arguments given in the preceeding paragraphs apply generically for any such a set. In fact, previous works have implicitly used this result when calculating entanglement between right- and left-movers before and after a Bogoliubov transformation in a quadratic system, see for example Refs. \cite{Ball_2006, GAO2007201}. By using the ladder operators before and after the transformation, one can construct different complete sets and so different partitions. 

Our point with the focus on the existence of charged superselection sectors is that the presence of conserved charges expands the number of possible sets that can be chosen and it does so in a very precise manner, as we have shown. When applying these ideas to field theoretical dualities, analysing the currents will allow us to quickly determine the mapping between momentum-space partitions, as any duality transformation must preserve the charged sectors of the original theory. 

\section{Application to the Bosonization duality}
\label{sec3}

The discussion in the previous Section was very general, so now we will give an explicit example in the form of fermionic QFTs in $1+1$ dimensions. Furthermore, we will show how the different momentum-space partitions of these systems are intimately related to the (abelian) Bosonization duality.

As a small reminder, the Bosonization duality allows us to map $(1+1)$d fermions into bosons with a precise connection between the Hamiltonians of each representation. At its heart, it relies on the identification of the fermionic density and current operators with derivatives of the canonically conjugate bosonic fields, i.e., $\partial_x\phi(x) = \rho(x) = :\psi^\dagger(x)\psi(x):$ and $\partial_x\pi(x) = j(x) = :\psi^\dagger(x)\gamma_5\psi(x):$, see for example, Refs. \cite{swieca, schulz1998fermi}.

From these basic equalities, a precise match between the operator contents of the bosonic and fermionic theories can be established and non-local expressions for the basic fields may be determined, as long as details explained in Ref. \cite{Headrick_2013} are properly taken into account (strictly speaking, the duality is only valid when mapping a compact scalar boson into a $\mathbb{Z}_2$-gauged fermion, though this will not affect our discussion here). Part of the power of the Bosonization duality stems from the wide range of QFTs to which it applies, from the famous Coleman correspondence between the sine-Gordon theory and the massive Thirring model \cite{coleman}, to extensions to Luttinger-Tomonaga liquid and charge-spin separation in condensed matter physics \cite{Haldane_1981, schulz1998fermi}, including obtaining band curvature corrections in real fermionic systems \cite{Haldane_1981, Pereira_2007}. Since all these applications follow from the basic current mappings, this Section's discussion on momentum-space entanglement will apply throughout this landscape of field theories.

\subsection{The Bosonization map and momentum modes}

The connection between the existence of multiple notions of momentum-space partitions in QFTs with superselection sectors and the Bosonization duality is quite straightforward. As previously mentioned, the linchpin of the duality transformation is the pair of operator identities $\partial_x\phi(x) = :\psi^\dagger(x)\psi(x):$ and $\partial_x\pi(x) = :\psi^\dagger(x)\gamma_5\psi(x):$. Therefore, since in momentum-space the derivative is simply a multipilation by $k$, the fermion density and bosonic field generate the same set (algebra) of physical observables in $1+1$ dimensions and we conclude, by taking the Fourier transform, that $\mathcal{H}^\phi_k=\mathcal{H}^\rho_k$ for all $k$. Thus, the effect of the duality in momentum-space is to map the ``density partition" of the fermionic theory into the ``$\phi$-field partition" of the bosonic representation and vice-versa. Although it is uncommon, we could also describe the bosonic observables in terms of the charged operators $:e^{\alpha\phi(x)}:$ (here the charge is the soliton number \cite{swieca}), with $\alpha = \frac{2\pi}{R}$ for period $R$ of the compact boson, and the associated tensor product partition $\bigotimes_k\mathcal{H}_k^\alpha$ would be equivalent to the fermion field decomposition $\bigotimes_k\mathcal{H}_k^\psi$.

Although the proof of the equality of tensor product factors was due to the fact that in this dimension we can use fermion field to reproduce canonical commutation relations via $[\rho(x),j(y)]=\partial_x\delta(x-y)$ with the appropriate filling of the Dirac sea \cite{lieb}, the mechanism of a duality transformation perfectly mapping different tensor product partitions into each other is what is expected to hold more generally, see Section \ref{sec5}. In fact, we may even draw a parallel with the conclusions at Section 2a of Ref. \cite{swieca}, that the key feature in bosonization is not the ``two-dimensional pathology" making the commutator of the fermion density and current be exactly the derivative of the canonical commutator of a boson, but rather the fact that the duality takes one momentum-space tensor product partition of one theory into an equivalent partition of another.

The abelian Bosonization scenario, in particular that of the Luttinger-Tomonaga liquid, also provides a simple demonstration of how the representations $\bigotimes_k\mathcal{H}_k^\psi$ and $\bigotimes_k\mathcal{H}_k^\rho= \bigotimes_k\mathcal{H}_k^\phi$ are physically distinct. As shown in Ref. \cite{lieb}, in $1+1$ dimensions, the ground state of a massless Dirac fermion with a four-fermion interaction $V(x-y)$ is of the form $\ket{\Omega}= e^{\hat{S}}\ket{0}$, with $\ket{0}$ being the free vacuum and $\hat{S}$ a bilinear on $\rho_k,j_k$. Thus, in the density representation this is a Gaussian state (in particular, separable in momentum-space), while in the fermion field representation it involves an exponential of a four-fermion term, thus having in general entanglement between momentum modes, as we will discuss next.

\subsection{Contrasting entanglement entropies}

We can go further in demonstrating the different natures of the distinct tensor product partitions by providing a few examples where the momentum-space entanglement entropy differ. The work of Refs. \cite{currents1, currents2} shows how one can formally write the Hamiltonian of a theory with conserved charges in terms of the density and current operators. In the particular case of a non-relativistic free theory, it becomes \cite{currents1},
\begin{equation}
\label{currentH}
    H = \frac{1}{8}\int d^dx \left[\nabla\rho(x)-2i\Vec{j}(x)\right] \rho^{-1}(x) \left[\nabla\rho(x)+2i\Vec{j}(x)\right].
\end{equation}

Therefore, a quadratic Hamiltonian which produces a Gaussian ground state in the field representation, becomes highly non-quadratic in the density representation. So in general we can expect that this leads to a state with nonzero entanglement.

From Eq. (\ref{currentH}) we see that even the simple free vacua can be difficult to study in the density representation. In order to give a more explicit example of the momentum-space entanglement changing with the representation, we can once again turn to the Bosonization duality.

The Coleman correspondence \cite{coleman} allows us to relate a mass perturbation of the Thirring model with a sine-Gordon perturbation of the massless boson and compare the momentum-space entanglement (in this paper, between fast and slow modes separated by the scale $\mu$) in the sine-Gordon and Thirring models at lowest nonzero order. And as we have seen, the first is equivalent to the density representation while the second is described in terms of the usual fermionic field.

We can quantitatively compare the different notions of entanglement via the path-integral method developed in Ref. \cite{paper1}. It works by taking the QFT's action functional $S[\phi]$ and integrating out the fast modes, i.e. the Fourier modes $\phi_{\Vec{k}}$ with $|\Vec{k}|>\mu$, following the Wilsonian RG procedure at finite temperature, which leads to an effective action $S_{\mu}^\beta$. Then, a replica technique is used, where the Rényi enttropies of integer order of the reduced density matrix are calculated via the following combination of partition functions:
\begin{equation}
\label{entropy}
    H_n(\rho_\mu) = \frac{1}{1-n}\lim_{\beta\to\infty}\left[\log Z_n(\mu,\beta) - n\log Z(\mu,\beta)\right],
\end{equation}
where the precise definition of the ``modified partition function" $Z_n(\mu,\beta)$ can be found in Ref. \cite{paper1}.

For the sine-gordon model with interaction written as $\alpha\cos(g\phi)$, the lowest-order terms in the effective action which contribute to the entropy (as shown in Ref. \cite{paper1}, those of order $\mathcal{O}(\alpha^2)$ which are non-local in Euclidean time) were derived in Ref. \cite{SVETITSKY1982423} and are given by,  
\begin{eqnarray}
    &&\frac{A_{\beta}^2(0)}{4}\left\{ \left[A_{\beta}^2(x-y,\tau-\tau')-1 \right] \cos g(\phi(x,\tau)+\phi(y,\tau')) \right.
\nonumber\\
    &+& \left.\left[A_{\beta}^{-2}(x-y,\tau-\tau')-1 \right] \cos g(\phi(x,\tau)-\phi(y,\tau')) \right\},
\end{eqnarray}
where $\phi(x,\tau)$ now only contains spatial Fourier modes in the interval $[-\mu,\mu]$, $A_{\beta}(x,\tau)=\exp[-\frac{g^2}{2}G_{\beta}(x,\tau)]$ and, 
\begin{equation}
    G_{\beta}(x,\tau) = \frac{1}{\beta}\sum_j\int_{|p|>\mu}\frac{dp}{2\pi}\frac{e^{ipx+i\omega_j\tau}}{p^2+\omega_j^2}.
\end{equation}

Then, putting the theory in a box of size $L$ with UV cutoff $\Lambda$, we perform the replica trick calculations as described in the Appendices of Ref. \cite{paper1}. First, we have, 
\begin{equation}
	A_{\beta\to\infty}^2(0)=\left(\frac{\mu}{\Lambda}\right)^{\frac{g^2}{2\pi}}.
\end{equation}
Next, the term from Eq. (\ref{entropy}) which leads to the momentum-space Rényi entropies at order $\mathcal{O}(\alpha^2)$ involves the expectation values of the trigonometric functions $\cos g(\phi(x,\tau)-\phi(y,\tau'))$ and $\cos g(\phi(x,\tau)+\phi(y,\tau'))$. Thus, we define, 
\begin{equation}
	G_{\beta, <}(x,\tau) = \frac{1}{\beta}\sum_j\int_{\frac{2\pi}{L}\leq|k|\leq\mu}\frac{dk}{2\pi}\frac{e^{ikx+i\omega_j\tau}}{k^2+\omega_j^2},
\end{equation}
and find the result,
\begin{widetext}
\begin{equation}
	\langle\cos g(\phi(x,\tau)\pm\phi(y,\tau'))\rangle = (\frac{2\pi}{L\mu})^{\frac{g^2}{2\pi}}\exp{\mp g^2G_{\beta, <}(x-y,\tau-\tau')},
\end{equation} 
with IR cutoff. 

To proceed, we expand all exponentials into series in $G_{\beta}(x,\tau)$ and $G_{\beta, <}(x,\tau)$ and use the Fourier sum $\frac{1}{\beta}\sum_j\frac{e^{-i\omega_j(\tau-\tau')}}{\omega_j^2+E^2}=\frac{e^{-E|\tau-\tau'|}}{2E}+\frac{\cosh(E|\tau-\tau'|)}{E(e^{\beta E}-1)}$ to identify which terms survive in the zero temperature limit. Although writing the full details of the final sum we need to calculate would be cumbersome, the general terms (suppressing integrals and factors of $e^{ip(x-y)}$ for a moment) will have the form
  \begin{eqnarray}
      \frac{g^{2(i+j)}}{i!j!}\int_0^\beta d\tau\int_0^\beta d\tau'\left[\frac{e^{-|p||\tau-\tau'|}}{2|p|}+\frac{\cosh(|p||\tau-\tau'|)}{|p|(e^{\beta |p|}-1)}\right]^i\left[\left(\frac{e^{-|k||\tau-\tau'|}}{2|k|}+\frac{\cosh(|k||\tau-\tau'|)}{|k|(e^{\beta |k|}-1)}\right)^j-\left(\frac{e^{-|k||\tau-\tau'|}}{2|k|}\right)^j\right].
  \end{eqnarray}

Where there are also $(-1)^{i}$ or $(-1)^{j}$ factors, depending on which cosine the contribution comes from.  Furthermore, the left-hand bracket comes from the expansion of $A_{\beta}^{-2}(x-y,\tau-\tau')$, while the right-hand one is a result of the exponential of $G_{\beta, <}(x-y,\tau-\tau')$ plus the usual manipulations and simplifications of our replica method, similarly to what happens in Appendix A of Ref. \cite{paper1}.

It is easy to show that the Euclidean time integrals are such that the only contributions surviving the $\beta\to\infty$ limit come exclusively from the hyperbolic cosine terms. Thus, re-introducing all suppressed factors, performing the integrals over $x$ and $y$ (which lead to the appearance of a delta function and the ``volume factor" $L$ usual for the momentum-space entanglement \cite{paper1, Balasubramanian:2011wt}), we find the Renyi entropy density
%
  \begin{eqnarray}
  \label{sinegordon}
        \frac{H_n}{L} &=& \frac{n\alpha^2}{n-1}\lim_{L\to\infty}\left(\frac{2\pi}{L\Lambda}\right)^{\frac{g^2}{2\pi}}\sum_{i+j \in 2\mathbb{N}}\frac{(-1)^i}{i!j!}\left(\frac{g^2}{4\pi}\right)^{i+j}  \int^*\frac{\prod_{a=1}^idp_a}{\prod_{a=1}^i|p_a|}\int^*\frac{\prod_{b=1}^jdk_b}{\prod_{b=1}^j|k_b|} \frac{2\pi\delta\left(\sum_{a=1}^ip_a+\sum_{b=1}^jk_b\right)}{\left(\sum_{a=1}^i|p_a|+\sum_{b=1}^j|k_b|\right)^2},
  \end{eqnarray}
such that $|p_a|\geq\mu$ and $\frac{2\pi}{L}\leq |k_b|<\mu$.

The exact expression is quite complicated, but note its key feature is that it involves an infinite series of momentum-space integrals. It may be also obtained from the Hamiltonian method developed in Ref. \cite{Balasubramanian:2011wt} if we write the interaction $\alpha\cos{g\phi(x)}$ as a series of even powers, regularize it and rearrange the permutations of momenta, which makes the factorials and signs match.

More importantly, we can now analyze the analogous entropy on the massive Thirring model side of the duality and see whether or not it agrees with Eq. (\ref{sinegordon}).

Using the expression derived in the Appendix C of Ref. \cite{Balasubramanian:2011wt}, the entanglement between same regions of momentum-space, but now in the field representation of the massive Thirring model, is proportional to the integral (over a specific set of momenta defined in the paper),
\begin{equation}
\label{fermion}
   \lambda^2\int^*\prod_{i=1}^4dp_i\frac{\delta(\sum_ip_i)(p_1p_3-m^2)(p_2p_4-m^2)}{\prod_i\sqrt{p_i^2+m^2}(\sum_i\sqrt{p_i^2+m^2})^2}.
\end{equation}

To compare to the sine-Gordon result, we expand it as a function of $m$ and use the Coleman correspondence \cite{coleman}, which makes $m$ proportional to $\alpha$.  

From Eq. (\ref{fermion}), $H_n(m)= H_n(0)+m^2H_n''(0)+\mathcal{O}(m^4)$. It turns out that $H_n(0)=0$, as in this limit the Thirring model is a scale-symmetric theory (see Ref. \cite{paper2} and the next Section for more details). The next term is
%
  \begin{eqnarray}
  H_n''(0) &\propto \int^*\prod_{i=1}^4dp_i  
       \delta(\sum_ip_i)\left\{\frac{2(p_1p_3+p_2p_4)}{\prod_i|p_i|(\sum_i|p_i|)^2}+\prod_i\sgn(p_i)\left[\frac{1}{(\sum_i|p_i|)^3}\left(\sum_i\frac{2}{|p_i|}\right)+\sum_j\frac{1}{|p_j|^2(\sum_i|p_i|^2)}\right]\right\}.
  \end{eqnarray}
\end{widetext}

Clearly it is structurally very different from Eq. (\ref{sinegordon}), meaning that even by identifying parameters under the duality, the momentum-space entanglement in the two representations is drastically different, which confirms our prediction. In particular, at the so-called ``free fermion point" of the Bosonization duality, when $g^2=4\pi$, the corresponding fermionic theory is non-interacting and thus its ground state is separable in momentum-space, while the density representation will still be entangled as can be seen from our nonzero result.

\section{Preservation of the separability of fixed points}
\label{sec4}

We gave examples in the previous Section of cases where the momentum-space entanglement in one representation is vanishing while being nonzero in another one. However, there is a scenario where all representations must have a separable ground state: when we deal with the entanglement between scales in RG fixed points.

In Ref. \cite{paper2} it was shown that RG fixed points have no entanglement between momenta at different scales. While the detailed derivation can be found in Ref. \cite{paper2}, the physical intuition is that a non-vanishing entanglement in momentum space would introduce a characteristic scale in the theory, thus violating the scaling symmetry underlying the RG fixed point.

The argument given relies only on the invariance of the ground state under the RG procedure and on the fact that at fixed points the field $\phi(x)$ is a scaling operator. Now, if we switch from the field representation to any other, say, to the density representation, the fact remains that at the fixed point the (irreducible) set of observables we are using to construct the physical operators of the theory will still be scaling fields. Therefore, the proof in Ref. \cite{paper2} follows once more and those QFTs have no entanglement between different momentum scales, regardless of the chosen partition. One can think of this intuitively as the fact that no matter the "coordinates" chosen, the scale-symmetry will still be preserved in the ground state, and so it must still be separable.

The Bosonization duality provides a great example to test this result by analyzing the (massless) Thirring model. In this limit it is an exactly solvable \cite{scarf, Klaiber:1967bmy} interacting theory of a single Dirac fermion in 1+1 dimensions, with Lagrangian density given by, 
\begin{equation}
	\mathcal{L} = i\Bar{\psi}\gamma^{\nu}\partial_\nu\psi + \frac{\lambda}{2}(\Bar{\psi}\psi)^2.
\end{equation}

Furthermore, it is well-known that this QFT is invariant under scaling transformations, which can be seen from the exact correlation functions derived in Ref. \cite{Klaiber:1967bmy}, and was pointed out by Wilson in Ref. \cite{opewilson}, where the fermionic field has scaling dimension,
\begin{equation}
	d_{\psi}= \frac{1}{2}+\frac{\frac{\lambda^2}{4\pi^2}}{1-\frac{\lambda^2}{4\pi^2}}.
\end{equation}
Hence, from the discussion at the beginning of this Section, we must be able to prove that the ground state of the Thirring model is separable with regards to the momentum scale partition in all representations.

This theory is dual to a massless free boson with Lagrangian $\mathcal{L} = \frac{1}{2g^2}\partial^\nu\phi\partial_\nu\phi$, with the relation $\frac{4\pi}{g^2}= 1+\frac{\lambda}{\pi}$ \cite{coleman, swieca}. Thus, for the bosonic/density representation, we trivially conclude that the ground state of the theory is the Fock vacuum and so has no entanglement between different scales.

As for the fermionic representation, the vanishing of the momentum-space entropy can be shown explicitly at lowest (nontrivial) order in the coupling $\lambda$ by analyzing Eq. (\ref{fermion}) in the $m=0$ limit,
\begin{equation}
\label{thirring}
S_{EE}(\mu) \propto \int^* \frac{\delta(\sum_ip_i)\sgn(p_1)\sgn(p_2)\sgn(p_3)\sgn(p_4)}{(|p_1|+|p_2|+|p_3|+|p_4|)^2},
\end{equation}
with the integral over momenta being such that there is always one momentum with modulus  smaller than $\mu$, another one with modulus always greater, and no set of momenta is repeated \cite{Balasubramanian:2011wt}, though the last condition may be ignored by dividing the result by the proper multiplicity factor.

Considering the form of the integrand and the region of integration is enough to prove that the expression above actually vanishes. Due to the presence of the sign functions, one only needs to show that to each set of momenta $\{p_1, p_2, p_3, p_4\}$ in the region, there is an associated set $\{p'_1, p'_2, p'_3, p'_4\}$ also being integrated over and such that an odd number of $p'_i$ change sign and $|p_1|+|p_2|+|p_3|+|p_4| = |p'_1|+|p'_2|+|p'_3|+|p'_4|$.

For instance, let, 
\begin{align}
    \begin{split}
        |p_1|&<\mu\\
        |p_2|&>\mu\\
        p_4&=-p_1-p_2-p_3.
    \end{split}
\end{align}
Then, to give an example of how cancellations will happen, consider the sets of momenta such that $p'_1=p_1 < 0$, $p_3>0$, $p'_3<0$ and $p_2, p'_2, p_4, p'_4$ are all positive. In this case, the contributions from sets $\{p_1, p_2, p_3, p_4\}$ and $\{p'_1, p'_2, p'_3, p'_4\}$ add to zero if we impose,
\begin{equation}
   p'_2-p'_3+p_1+p'_2+p'_3=p_2+p_3+p_1+p_2+p_3,
\end{equation}
which means, $p'_2 = p_2+p_3$, and we can choose $p'_3 = -p_3$. By definition, $p_2,p_3>0$, and therefore, $p'_2 > \mu$, so that the set found is still in the region of integration.
Furthermore, the procedure is always valid in a sufficiently small neighborhood around $(p_1, p_2, p_3, p_4)$, so the cancellation occurs for the integral over a volume as well.

From this, we proceed by defining similar rules associating different regions of momenta to each other in such a way as to always have a vanishing net integral until we cover all possibilities. Uniqueness of pairs is imposed by construction, since the equality of sums of absolute values has enough solutions for more conditions to be applied. And so, we conclude that the full integral of Eq. (\ref{thirring}) vanishes as demanded by scale-symmetry.

The importance of having an RG fixed point in this example is made clear if we try to apply the same method to the models studied in Ref. \cite{lieb}, where the four-fermion vertex is not constant. In the bosonic representation these are free theories without scale invariance. Our construction would no longer work, as the integrals would involve the momentum-dependent vertex and the cancellations could not be guaranteed for every point in the region of integration.

Returning to the Thirring model, we could expect there to be a proof to all orders in perturbation theory following from our method in Ref. \cite{paper1}, but we leave such an investigation for future work. Note, however, that another derivation of the separability of this QFT's ground state was shown in \cite{mcdermott}, via a scaling argument of a different nature than ours in Ref. \cite{paper2}.
 
\section{General and local dualities}
\label{sec5}

For this final Section, we aim to answer two questions: first, why does the Bosonization duality, as proven in Ref. \cite{Headrick_2013}, preserve the real space entanglement of a region, while the Wannier duality for the Ising and other lattice models does not \cite{radivcevic2016entanglement}? Secondly, what does this distinction imply about the mapping of different momentum-space tensor product partitions in a duality transformation?

In order to address these topics, we will use the language of von Neumann algebras somewhat more than in the rest of this paper. This will allow us to answer the questions posed in a rather straightforward manner and understand more the structures involved in duality transformations in general.

To move forward, me must understand how exactly a duality transformation acts on the algebras of observables of a QFT. The fundamental insight of Ref. \cite{bondalgebra} is that the most general characteristic of any duality is that it is an isomorphism of the von Neumann bond algebras, the algebra of operators associated with the terms of a local Hamiltonian (see Section 3, and in particular subsection 3.3, of Ref. \cite{bondalgebra} for a more detailed explanation). Because we are focusing on local Hamiltonians, the bond algebras of a region of real space (for example, a number of sites in the Ising model) is automatically a proper subset of the maximal algebra of observables associated with the same region, and it is from this fact that we will be able to answer the questions posed.

First, from the characterization of dualities as isomorphisms of bond algebras plus the well-known mathematical result that any von Neumann algebra isomorphism $\Phi$ acts on a operator $\mathbb{O}$ as $\Phi(\mathbb{O}) = U\left(\mathbb{O}\otimes1_{\mathcal{H}}\right)U^\dagger$ for appropriate Hilbert space $\mathcal{H}$ and unitary $U$ (see Theorem 3.3 of Ref. \cite{bondalgebra}), the results of Ref. \cite{radivcevic2016entanglement} immediately follow: the algebras of observables whose ground state entanglement entropies were studied are precisely the algebras mapped into each other by the Wannier duality, in other words, the bond algebras themselves. Additionally, the fact that generically a von Neumann algebra isomorphism involves a tensoring with an auxiliary Hilbert space explains why there is a change in the representation of the bond algebras. 

Furthermore, this also explains why in the cases studied in Ref. \cite{radivcevic2016entanglement} fixing the eigenvalue of an operator (such as an edge spin) is at times required in order to make the representations on both sides of the duality have same dimension, and thus having entropies which can be meaningfully compared. Since a general isomorphism may involve tensoring with an auxiliary Hilbert space, we must restrict its image to a subspace (i.e., a diagonal block of the new density matrix), so that it becomes a unitary, thus preserving the entropy. Through the bond algebraic formalism of Ref. \cite{bondalgebra}, therefore, the results of Ref. \cite{radivcevic2016entanglement} are generalized to all duality transformations involving local Hamiltonians.

Now, for many dualities, such as Wannier, the bond algebras are the largest structure preserved by the transformation, i.e., larger local algebras are not isomorphic. This is easily seen in the Ising model, where the dual variables are non-local observables on the original spins, since they are associated to bonds/links instead of sites. Therefore, we conclude that generically a duality will not preserve the full real space entanglement of a region, since the minimal requirement for such a transformation is only the isomorphism of the bond algebras. This implies that the result for the abelian Bosonization obtained in Ref. \cite{Headrick_2013} is the outlier which must be understood: in Section \ref{sec3} we pointed out how this specific duality derives from the equalities $\partial_x\phi(x) = \rho(x) = :\psi^\dagger(x)\psi(x):$ and $\partial_x\pi(x) = j(x) = :\psi^\dagger(x)\gamma_5\psi(x):$, but since the density and current operators generate the full set of physical operators of the theory, this implies that the Bosonization duality is an isomorphism of the full local algebras of observables, which thus generalizes the previous result to the complete entanglement entropy associated with a region in real space.

Such conclusion leads us to define another, stronger (more restrictive) notion of duality mapping. While the most general dualities are isomorphisms of only the bond algebras, a subset of those can be called ``local dualities", where these isomorphisms extend to all the local algebra. It is then for this subset that we can expect the preservation of the entanglement entropy in all cases. 

Interestingly, for local dualities of continuum theories, since the local algebra of observables is of Type III as shown in Ref. \cite{arakitype3}, the isomorphism is guaranteed to always be a unitary, see Ref. \cite{unitarytype3}, and so the entropy must be exactly the same on both sides of the duality, without need for the specification of the eigenvalue of some operator.

It is for a ``local duality" that our discussion in Section \ref{sec3} on the mapping between different momentum-space partitions applies: as we have seen, the definition of these tensor product structures relies on having maximal sets of commuting operators in order to characterize the degrees of freedom and thus the momentum modes. Therefore, in order for different partitions to be mapped into each other, this requires an isomorphism of these maximal abelian von Neumann algebras which thus extends to an isomorphism of the full local algebra (formally one adjoints the ``canonical momenta" to the algebra, whose commutation relations will also be preserved).
With this, we both conclude that similar results as in Section \ref{sec3} apply to all local dualities and that such statements will generally only be valid in such cases. The question remains as to what happens if we analyze the momentum-space entanglement for a general duality. We will not do so here, in particular because the notion of ``bond locality" that characterizes the bond algebras is no longer valid in momentum space (where interactions are highly non-local), but we leave it for further work.

\section{Conclusions}

We studied how the momentum-space entanglement structure of a QFT is changed in the presence of conserved charges. Namely, new ways of partitioning the Hilbert space into a tensor product of momentum modes are introduced in a controlled way, through the fact that the currents can be used to construct other complete set of commuting operators. We have also seen what this implies for duality transformations which completely preserve the local algebra of observables: different tensor product structures are mapped into each other and the momentum-space entanglement between modes is not guaranteed to be preserved (except for RG fixed points), as we showed explicitly for the abelian Bosonization duality in $1+1$ dimensions.

Our findings point to the importance of the choice of a maximal set of commuting observables in defining what is meant by the ``momentum-space operators" of a field theory. While this means that there is no unique notion (up to a choice between maximality or additivity, see Ref. \cite{Casini:2020rgj}) as in the real space case, it also allows one to choose which notion of momentum modes is more practical for a given context. In particular, in Ref. \cite{paper2} it was briefly discussed how a momentum-space representation could be useful to the development of new numerical methods meant to study ground states of critical QFTs as well as how these notions could be mathematically defined through the theory of von Neumann algebras. The existence of multiple partitions then allows us to choose whichever is more convenient for these purposes. 

There are a number of ways to take this work forward. In particular, as we mentioned in Section \ref{sec2}, we only studied the presence of ``ordinary" symmetries and so an interesting direction would be to do a similar analysis for theories with gauge or generalized symmetries. 

Real space entanglement in gauge QFTs is well understood, as detailed in Refs. \cite{estring, nonab, soni, lin}, while there is a growing interest for the relation between quantum information and generalized symmetries, see Refs. \cite{Casini:2020rgj, PhysRevD.109.105026}, since they can be used to describe topological phases \cite{yoshida, LUO20241} and these have a rich entanglement structure \cite{Kitaev:2005dm, XG}. Both topics have yet to be approached in momentum space and the gauge theory case may involve more complications, since gauge ``symmetry", specially for nonabelian groups, forbids us to cleanly associate momentum-space observables to field operators as we have been doing so far and there are no proofs as in Refs. \cite{currents1, currents2} that the currents form an irreducible set of operators. 

Nevertheless, given the strong constraints both gauge and generalized symmetries impose on the RG flow of a QFT via anomalies \cite{shao2024whatsundonetasilectures} and the RG's inherent connection to entanglement between momentum modes \cite{paper2}, should a way to properly formulate the problem be found, they would certainly be interesting results.

\begin{acknowledgments}
	We thank Zohar Nussinov and Zack Weinstein for useful comments and discussions. 
	We acknowledge financial support by the Deutsche Forschungsgemeinschaft (DFG, German Research Foundation), through SFB 1143 project A5 and the W{\"u}rzburg-Dresden Cluster of Excellence on Complexity and Topology in Quantum Matter-ct.qmat (EXC 2147, Project Id No. 390858490). 
\end{acknowledgments}

\bibliography{citations}

\end{document}